\documentclass[aps,prd,twocolumn,nofootinbib,unsortedaddress]{revtex4-2}

\newcommand\prx{PRX}
\usepackage{aas_macros}
\usepackage{ns_bimodal}
\usepackage{analytic_scaling_RMS}

\usepackage{hyperref}

\usepackage{amsmath}
\usepackage{amsfonts}
\usepackage{amssymb}
\usepackage{mathtools}
\usepackage{mathrsfs}
\usepackage{bm}
\usepackage{graphicx}
\usepackage{xspace}

\newcommand{\param}{{\boldmath\lambda}}
\newcommand{\Param}{X}
\newcommand\MedianOf[1]{{#1}_{0.5}}

\usepackage{xcolor}
\newcommand\editremark[1]{{\color{red} #1}}
\newcommand\SkipPreliminary[1]{}
\newcommand\SkipCrapAtEnd[1]{}

\definecolor{todocomment}{HTML}{D62728}
\definecolor{dwcomment}{HTML}{1F77B4}
\definecolor{rocomment}{HTML}{9467BD}
\definecolor{lwcomment}{HTML}{FF7F0E}
\definecolor{jrcomment}{HTML}{2CA02C}

\newcommand{\Resum}{\textsc{TEOBResumS}\xspace}

\newcommand{\IMRPDTv}{\textsc{IMRPhenomD\_NRTidalv2}\xspace}

\newcommand\E[1]{{\left\langle #1 \right \rangle}}
\newcommand\unit[1]{\mathrm{#1}}
\newcommand\rate{\mathcal{R}}

\newcommand\mchirp{{\mathcal{M}_{\text{c}}}}
\newcommand\mc{\mchirp}

\newcommand\Msun{{\mathrm{M}_{\odot}}}

\newcommand\cqg{CQG}

\usepackage{data_macros}

\renewcommand{\BibitemShut}[1]{}

\begin{document}

\title{
  Effects of waveform systematics on inferences of neutron star population properties 
  and the nuclear equation of state
}


\author{Anjali B. Yelikar}
\email{ay2016@rit.edu}
\affiliation{Rochester Institute of Technology, Rochester, New York 14623, USA}
\affiliation{Vanderbilt University, Nashville, TN 37235, USA}

\author{Richard O'Shaughnessy}
\affiliation{Rochester Institute of Technology, Rochester, New York 14623, USA}

\author{Daniel Wysocki}
\affiliation{University of Wisconsin-Milwaukee, Milwaukee, WI 53201, USA}

\author{Leslie Wade}
\affiliation{Kenyon College, Gambier, Ohio 43022, USA}


\date{\today}

\begin{abstract}
Gravitational waves from inspiralling neutron stars carry information about matter at extreme gravity and density. The binary neutron star (BNS) event GW170817 provided, for the first time, insight into dense matter through this window. Since then, another BNS (GW190425) and several neutron star-black hole events have been detected, although the tidal measurements were not expected to be well-constrained from them. Collective information regarding the behavior of nuclear matter at extreme densities can be done by performing a joint population inference for the masses, spins, and equation-of-state~\cite{Wysocki:2020myz} to enable better understanding. This population inference, in turn, relies on accurate estimates of intrinsic parameters of individual events. In this study, we investigate how the differences in parameter inference of BNS events using different waveform models can affect the eventual inference of the nuclear equation-of-state. We use the state-of-the-art model \Resum with \IMRPDTv as a comparison model. 
\end{abstract}

\maketitle

\section{Introduction}
\label{sec:intro}


Neutron stars have proved to be excellent laboratories for understanding the behavior of cold nuclear matter, be it observations in the electromagnetic spectrum
or gravitational waves. The ground-based gravitational wave detectors, LIGO-Virgo-KAGRA (LVK)~\cite{2015CQGra..32g4001T,VIRGO:2014yos,KAGRA:2020tym}, have observed about 90
signals from binary mergers till March 2020 since the first detection in September 2015~\cite{LIGO-O2-Catalog,LIGO-O3-O3a-catalog,LIGO-O3-O3a_final-catalog,LIGO-O3-O3bcatalog}. The most recent population paper from the LVK collaboration paper~\cite{LIGO-O3-O3bpop} estimated the
merger rate of binary neutron stars to be between 10 and 1700 Gpc$^{-3}$yr$^{-1}$. To date, only two confirmed BNS have
been found in the gravitational wave domain- GW170817~\cite{LIGO-GW170817-bns} and
GW190425~\cite{Abbott:2020uma}. Gravitational wave measurements from GW170817 alone were able to place constraints on
existing EOS models
~\cite{LIGOScientific:2019eut,LIGOScientific:2018cki,Annala:2017llu,Most:2018hfd,Shibata:2019ctb,Raithel:2018ncd}. The
observations from its electromagnetic counterpart, AT2017gfo, crucial in identifying the sky localization and distance
with certainty, also provided additional information from which to understand the interior of neutron stars
better~\cite{Dietrich:2020efo,Coughlin:2018miv,Coughlin:2018fis,Margalit:2017dij,Raaijmakers:2019dks}.
Direct observations of pulsars alone in various regions of the EM spectrum and results from the NICER experiment have provided insight 
regarding the interior of neutron stars, helping constrain their mass-radius relationship and in turn helping understand which of the equation of state models accurately predict the results
from observed data~\cite{Ozel:2010fw,Steiner:2010fz,Lattimer:2013hma,Watts:2016uzu,Miller:2019cac}. Joint constraints from GW and EM observations have also been 
insightful in providing better constraints on existing EoS models~\cite{Kedia:2024hvw}.

  

Gravitational waves from neutron stars also provide support for the existence of higher mass NS in contrast to the neutron stars observed in Galactic BNS. We expect to see many more BNSs with improving sensitivity of current and next generation GW detectors~\cite{2020LRR....23....3A}, and combining information from future detections will enable better constraints on the pressure-density behavior of high-density matter, hence providing an understanding and validity of existing equation-of-state models from gravitational wave information alone. In a follow-up paper on the discovery of GW170817~\cite{LIGO-GW170817-EOS}, information from gravitational waves was used to constrain the neutron star radius and the nuclear equation of state. 
\\

In Wysocki et al. ~\cite{Wysocki:2020myz}, simultaneous measurement of mass and spin distributions along with the equation of state (EoS) was demonstrated with a set of fiducial binary neutron star merger signals. The study used the Bayesian population inference code, \textsc{PopModels}~\cite{popmodels}, and a spectral representation for the equation of state introduced by Lindblom~\cite{Lindblom:2010bb}. Based on this work, we propose to investigate the effect of waveform systematics on the joint inference of NS population properties and the EoS. Similar hierarchical inference techniques involving stacking gravitational wave information from binary neutron stars have been described in ~\cite{Ray:2022hzg,Golomb:2021tll}.

This paper is organized as follows.
In Section ~\ref{sec:methods}, we review the gravitational wave parameter inference tool \textsc{RIFT} and population inference tool \textsc{PopModels}; the source population model and waveform models assumed. We describe one fiducial set of synthetic sources populated using astrophysically motivated models. In Section \ref{sec:results}, we present
comparisons of the performance of PE using different waveform models and quantify these differences with techniques also used in ~\cite{Yelikar:2024wzm}. We further 
investigate the equation of state results inferred from a joint population and EoS analyses. We show that waveform modeling uncertainties will be significant when
combining information from multiple detections to infer the properties of a neutron star. In Section \ref{sec:discussion}, we evaluate our results within the context of 
existing literature. In Section \ref{sec:conclude}, we summarize our results and discuss their potential applications to future detections of BNS sources and improvements in
our current setup, such as combining information from neutron star-black hole (NSBH) sources. 


\section{Methods}
\label{sec:methods}

\subsection{\textsc{RIFT} review}
A coalescing compact binary in a quasi-circular orbit can be entirely characterized by its intrinsic
and extrinsic parameters.  By intrinsic parameters, we refer to the binary's detector-frame masses $m_i$, spins $\chi_{i}$, and any quantities
characterizing matter in the system, $\Lambda_i$. By extrinsic parameters, we refer to the seven numbers needed to 
characterize its spacetime location and orientation; luminosity distance ($d_{L}$), right ascension ($\alpha$), declination ($\delta$), inclination ($\iota$), polarization ($\psi$), coalescence phase ($\phi_{c}$), and time ($t_{c}$).
  We will express masses in solar mass units and
 dimensionless nonprecessing spins in terms of cartesian components aligned with the orbital angular momentum
 $\chi_{i,z}$, as we use waveform models that do not account for precession. We will use \bm{$\lambda$} and $\theta$ to
refer to intrinsic and extrinsic parameters. 

$$ \bm{\lambda} : (\mathcal{M},q,\chi_{1,z},\chi_{2,z},\Lambda_{1},\Lambda_{2}) $$
$$ \theta : (d_{L},\alpha,\delta,\iota,\psi,\phi_{c},t_{c}) $$

\textsc{RIFT} \cite{gwastro-PENR-RIFT,gwastro-RIFT-Update}
consists of a two-stage iterative process to interpret gravitational wave data $d$ via comparison to
predicted gravitational wave signals $h(\lambda, \theta)$. In one stage, for each  $\lambda_\beta$ from some proposed
``grid'' $\beta=1,2,\ldots N$ of candidate parameters, \textsc{RIFT} computes a marginal likelihood 
\begin{equation}
 {\cal L}_{\rm marg}\equiv\int  {\cal L}(\bm{\lambda} ,\theta )p(\theta )d\theta
\end{equation}
from the likelihood ${\cal L}(\bm{\lambda} ,\theta ) $ of the gravitational wave signal in the multi-detector network,
accounting for detector response; see the \textsc{RIFT} paper for a more detailed specification~\cite{gwastro-PENR-RIFT,gwastro-RIFT-Update}.  
In the second stage,  \textsc{RIFT} performs two tasks. First, it generates an approximation to ${\cal L}(\lambda)$ based on its
accumulated archived knowledge of marginal likelihood evaluations 
$(\lambda_\beta,{\cal L}_\beta)$. Gaussian processes, random forests, or other
suitable approximation techniques can generate this approximation. Second, using this approximation, it generates the (detector-frame) posterior distribution
\begin{equation}
\label{eq:post}
p_{\rm post}=\frac{{\cal L}_{\rm marg}(\bm{\lambda} )p(\bm{\lambda})}{\int d\bm{\lambda} {\cal L}_{\rm marg}(\bm{\lambda} ) p(\bm{\lambda} )}.
\end{equation}
where prior $p(\bm{\lambda})$ is prior on intrinsic parameters like mass and spin.    The posterior is produced by
performing a Monte Carlo integral:  the evaluation points and weights in that integral are weighted posterior samples,
which are fairly resampled to generate conventional independent, identically distributed ``posterior samples.''
For further details on RIFT's technical underpinnings and performance,   see
\cite{gwastro-PENR-RIFT,gwastro-RIFT-Update,gwastro-PENR-RIFT-GPU,gwastro-mergers-nr-LangePhD}.

To facilitate robust sampling of the most a priori important regions, particularly in NS tidal deformability $\Lambda_i$, \textsc{RIFT} further adopts a hierarchical strategy,
employing targeted sampling priors for early iterations to guarantee thorough coverage.   As described in prior work
\cite{gwastro-RIFT-Update}, the initial range of $\Lambda_i$ used to construct the first set of likelihood training data
uses an interval in $\Lambda(m)$ between $\text{min}[50,0.2\Lambda_{fid}(m)]$ and $\text{max}(150,2\Lambda_{\rm
  fid}(m))$, where $\Lambda_{\rm fid} = 3000((2.2-m/M_\odot)/1.2))^2$.  This choice guarantees comprehensive sampling
over the range of $\Lambda$, most likely to be consistent with physically plausible EoS.   To cover the
low-$\Lambda$ region further, the next few iterations adopt a nonuniform triangular prior in $\Lambda_i$, which goes to zero at
the maximum allowed value.  After these first few iterations establish seed training data, subsequent iterations are
repeated using the target inference priors, until the posterior converges.

\subsection{Waveform models and analysis settings}
The tidal waveform models used in this study are \IMRPDTv and \Resum. NRTidalv2 models~\cite{Dietrich:2019kaq} 
are improved versions of NRTidal~\cite{Dietrich:2017aum} models, which 
are closed-form tidal approximants for binary neutron star coalescence and have been analytically added to selected 
binary black hole GW model to obtain a binary neutron star waveform, either in the time or frequency domain. 
\Resum~\cite{2018PhRvD..98j4052N}  is another but unique time-domain EOB formalism that includes tidal effects for all modes $\ell \leq 4$, but no $\mathit{m}=0$ and 
models tidal effect up to $\Lambda_{1,2} < 5000$ and for spins up to 0.5.

For parameter inference, we use  \Resum for injections and recovery and \IMRPDTv for the recovery, starting the 
signal evolution and likelihood integration at $20\unit{Hz}$, performing all analysis with $4096\unit{Hz}$ time series in Gaussian noise with
known advanced LIGO design PSDs \cite{LIGO-aLIGODesign-Sensitivity-Updated}.  The BNS signal is generated for 
4096 seconds but analysis was performed only on 256 seconds of data. 
For each synthetic event and interferometer, we use the same noise realization for all waveform 
approximations. Therefore, the differences between them arise solely due to waveform systematics.
\Resum and \IMRPDTv approximants are used with $\ell=4$ and $\ell=2$ settings respectively.

\subsection{\textsc{PopModels} review}

To jointly infer both the nuclear EoS and NS-NS population, we use the framework introduced in two previous works led by
Wysocki \cite{Wysocki:2020myz,gwastro-PopulationReconstruct-Parametric-Wysocki2018}, which we denote as ``bayesian
parametric models'' (henceforth denoted as BPM).   Within this framework, binaries with intrinsic parameters $\param$ merge at
a rate ${\cal R}p(\param)$.   Generally the merger rate ${\cal R}$ and population distribution $p$ depend on population
and EoS hyperparameters $\Param$.  BPM's inputs are (a) the underlying parameterized population model $p(\param|\Param),
{\cal R}(\Param)$ and (b) an estimate $\mu(\Param)$ of how many events a given experiment should find on average.
Following BPM, we estimate this rate using a characteristic sensitive 4-volume denoted $VT(\param)$.
%
%
In terms of these ingredients, BPM expresses the likelihood of an astrophysical BBH population with parameters $\Param$
as a conventional inhomogeneous Poisson process:
\begin{equation}
  \mathcal{L}(\rate, \Param) \propto
  e^{-\mu(\rate, \Param)}
  \prod_{n=1}^N
    \int \mathrm{d}\param \, \ell_n(\param) \, \rate \, p(\param\mid\Param),
  \label{eq:inhomog-poisson-likelihood}
\end{equation}
where $\mu(\rate,\Param)$ is the expected number of detections under a given population parameterization $\Param$ with
overall rate $\rate$ and where $\ell_n(x)=p(d_n|x)$ is the likelihood of data $d_n$---corresponding to the $n$th detection---given binary parameters
$x$.
Like previous EoS inferences within this framework \cite{Wysocki:2020myz}, all single-event marginal likelihoods
$\ell_n$ are provided by interpolating the training likelihood data provided by conventional RIFT single-event inference.

We employ precisely the same NS population model proposed in previous work with BPM on EoS \cite{Wysocki:2020myz}.
Specifically,  motivated by observations of galactic binary neutron stars \cite{2016ARA&A..54..401O,2018MNRAS.478.1377A}, we employ a two-component population of neutron stars, with overall minimum and maximum masses set by the nuclear equation of state:
\begin{gather}
  p(x) \propto
  \sum_{k=1}^K w_k \,
    \mathscr{M}_k(m_1) \, \mathscr{M}_k(m_2) \,
    \mathscr{S}_k(\chi_{1,z}) \, \mathscr{S}_k(\chi_{2,z})
  \label{eq:pop-model}
\end{gather}
everywhere that $m_{\mathrm{min}}(\boldsymbol{\gamma}) \leq m_2 \leq m_2 \leq m_{\mathrm{max}}(\boldsymbol{\gamma})$ and
$\Lambda_i = \Lambda(m_i, \boldsymbol{\gamma})$, and zero elsewhere.  $\mathscr{M}_k$ and $\mathscr{S}_k$ represent the
mass and spin distributions for the $k$th sub-population, respectively.  We model the $\mathscr{M}_k$'s as Gaussians
with unknown mean and variance.  The $\mathscr{S}_k$'s are assumed to follow beta distributions bounded by $|\chi_z| <
0.05$, again with unknown mean and variance.  For simplicity's sake, we assert that the means and variances don't change
between the primary and secondary NS.  The $\Lambda_i = \Lambda(m_i, \boldsymbol{\gamma})$ constraint introduces delta
functions into the expression for $p(x)$

\subsection{Equation of state representation}

In this work we adopt the spectral EoS parameterization introduced by Lindblom  \cite{2010PhRvD..82j3011L}, implemented
in Carney et al \cite{2018PhRvD..98f3004C} in \textsc{LALSuite} \cite{lalsuite},
and previously used to interpret GW170817 \cite{LIGO-GW170817-EOS}.  
In this specific representation, the nuclear equation of state relating energy density $\epsilon$ and pressure $p$ is
characterized by a low-density SLy EoS joined to a spectral representation  
 at $p_0=5.4\times 10^{32}\unit{dyne}\; \unit{cm}^{-2}$, using a high-density 
four-parameter spectral model characterized by its adiabatic index $\Gamma(p)$:
\begin{eqnarray}
\ln \Gamma(p) = \sum_{k=0}^3 \gamma_k [\ln (p/p_0)]^k,
\end{eqnarray}
where $\gamma_k$ are expansion coefficients.

We employ the already-deployed \textsc{PopModels} interface to the underlying lalsuite implementation of this equation of state.
As a fiducial EoS, we will adopt the spectral approximation to APR4 from \textcite{2010PhRvD..82j3011L},
 given by $\gamma_0 = 0.8651$, $\gamma_1 = 0.1548$, $\gamma_2 = -0.0151$, $\gamma_3 = -0.0002$.
 All of our analyses use uninformative uniform priors on the spectral EoS parameters, following previous work
\cite{2018PhRvD..98f3004C,LIGO-GW170817-EOS,Wysocki:2020myz}; $\gamma_0$:  U$[0.2, 2]$, $\gamma_1$: U$[-1.6, +1.7]$, 
$\gamma_2$: U$[-0.6, +0.6]$,  $\gamma_3$: U$[-0.02, +0.02]$.  


The pressure and nuclear density of the neutron star's interior are modeled by spectral parameterization.

\subsection{Fiducial injections and Source population model}


\begin{figure}
\includegraphics[scale=0.45]{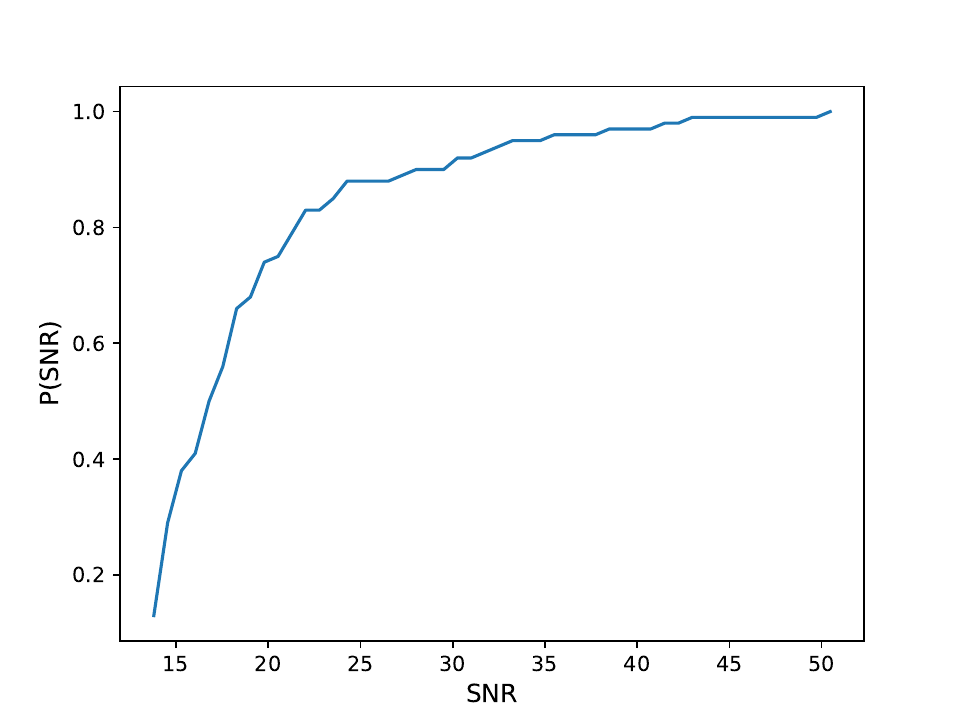}
\caption{Cumulative SNR distribution for a synthetic population of 100 events each drawn from the fiducial BNS populations described in Sec. II C. To avoid ambiguity, this figure shows the expected SNR (i.e., the SNR evaluated using a zero-noise realization).}
\label{zero-noise-snr}
\end{figure}

We consider one universe of 100 synthetic signals for a
2-detector network (HL), with signals populated from the following model. 
For our fiducial model, we took a two-component (K = 2 in Eq. ) mass distribution based on the galactic neutron 
star constraints from \textcite{2018MNRAS.478.1377A}, in the first row of Table 3. Rather than take their maximum a posteriori 
values, we approximated their reported estimates as Gaussians, and took a single draw, resulting in 
$\mathbb{E}[m]_1 = 1.34 \, M_\odot$, $\mathrm{Std}[m]_1 = 0.05 \, M_\odot$, $\mathbb{E}[m]_2 = 1.88
\, M_\odot$, $\mathrm{Std}[m]_2 = 0.32 \, M_\odot$
, and relative weights of $6:4$. 
For the low mass component’s spin distribution, we utilize a zero-spin model, attainable using a $\beta$ distribution 
with $\mathbb{E}[\chi_z]_1 = 0$ and
$\textrm{Std}[\chi_z]_1 \to 0$. For the high mass component, however, we expect higher spins, as these 
would likely be recycled pulsars \cite{2008LRR....11....8L}, and so we adopt fiducial choices $\mathbb{E}[\chi_z]_2 = 0.02$ and $\mathrm{Std}[\chi_z]_2 = 0.01$.

%

Our population analyses use uninformative priors, 
uniform (U) in $\mathbb{E}[m]$, $\mathbb{E}[\chi_z]$, $\mathrm{Std}[m]$, and log-uniform (LU) in each sub-population's rate 
and $\mathrm{Std}[\chi_z]$; $\mathcal{R}$ [$\mathrm{Gpc}^{-3} \, \mathrm{yr}^{-1}$]: LU $[1, 10^5]$, $\mathbb{E}[m]$ [$M_\odot$]: U $[0.9, 2.9]$, 
$\mathrm{Std}[m]$ [$M_\odot$]: LU $[0.05, 5]$, $\mathbb{E}[\chi]$: U $[-0.05, +0.05]$ and $\mathrm{Std}[\chi]$: U $[0, 0.05]$. 
To account for observations of galactic neutron stars, rather than reanalyze all galactic observations ourselves, we
employ pre-digested prior constraints on this same two-component mass model provided by Table 3 of
\cite{2018MNRAS.478.1377A}.
In particular, we use an (improper) prior in the maximum neutron star mass $m_{\rm max}(\boldsymbol{\gamma})$,
extending from $1.97 M_\odot$ to infinity, to account
for the impact of the most recent well-determined NS masses on the inferred NS maximum mass  \cite{2013Sci...340..448A,2016arXiv160501665A,2019arXiv190406759C}.

Figure~\ref{zero-noise-snr} shows the cumulative SNR distribution (under a ``zero-noise" assumption) 
of the specific synthetic population generated from this distribution. Compared to GW170817's confident detection, 
which was a BNS merger that occurred at $40\unit{Mpc}$ detected by LIGO-Virgo with a SNR of 32.4, 
the majority of the signals in this fiducial population have SNRs below or near the typical detection criteria for a BNS merger, 
with some having high enough amplitudes. 

\section{Results}
\label{sec:results}

In our calculations, we first construct a single large synthetic sample of merging binary NS drawn from our population
model, weighted by their probability of detection.  In this work, we have constructed our synthetic population using the
TEOBResumS waveform model.  Rather than regenerate many small independent synthetic universes,
we then conceive of experiments in which different subsets of our full sample are identified via a GW survey.   Thus,
for each member of our large synthetic sample, we have performed full source parameter inference with \textsc{RIFT} in an
EoS-agnostic phenomenological form, providing the underlying training data $\{\lambda_\alpha,\ln {\cal L}_{\rm
  marg}(\lambda_\alpha)\}$ used to interpolate the \textsc{RIFT} log-likelihod over $\lambda$.

\subsection{Single event parameter inference}
Any systematic effect of source model waveforms on EoS and population inference must be inherited from source parameter
inference.   In this section, we briefly review the similarities and differences between different interpretations of
synthetic events, depending on the waveform model used.

\begin{figure}
\includegraphics[scale=0.45]{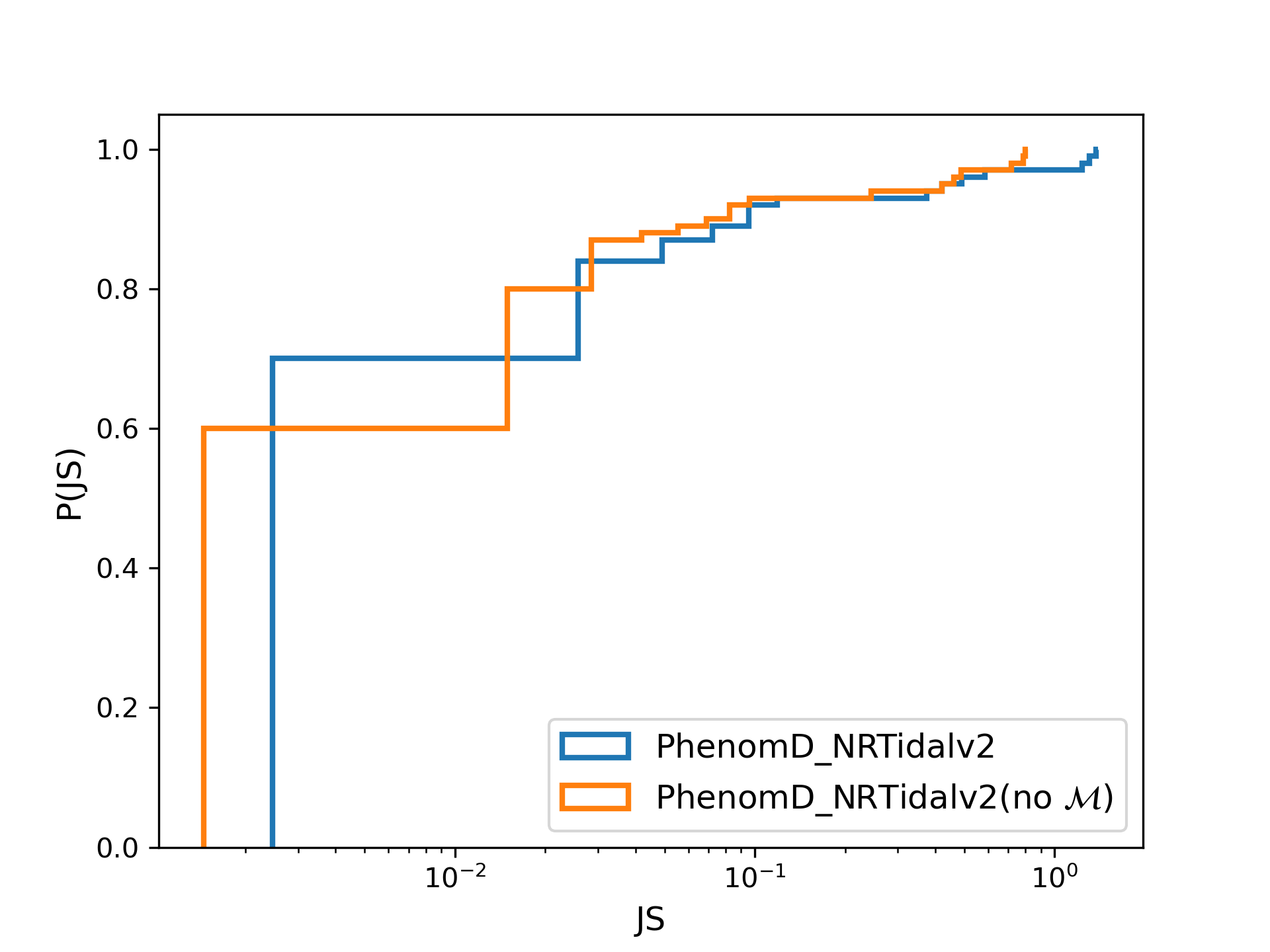}
\includegraphics[scale=0.45]{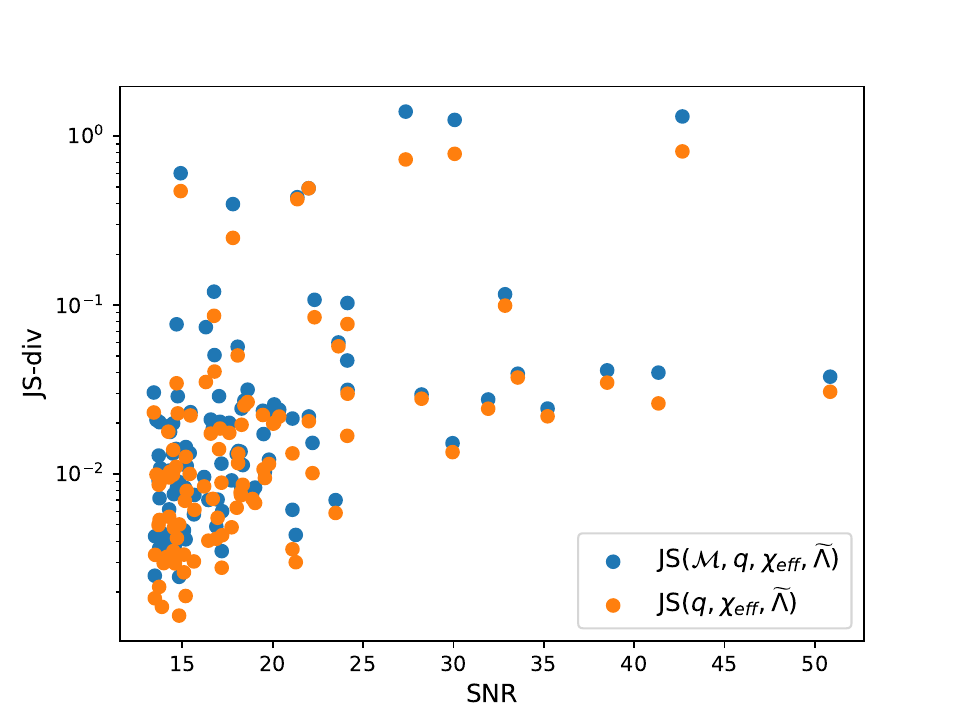}
\caption{\emph{Top panel}: JS-divergence values of parameters $\mc$, q,  $\chi_{eff}$ and
  $\widetilde{\Lambda}$) distribution for analysis on \IMRPDTv with \Resum-lmax4 injections.
\emph{Bottom panel}: Scatter plot showing how the zero-noise SNR of the injections from the population relates to the JS divergence value between 
\Resum and \IMRPDTv samples for the parameters noted in the legend.}
\label{JS-div}
\end{figure}

We quantify the differences between the various parameter estimation analyses using Jensen-Shannon (JS) divergences. 
The divergence values shown are the summation of individual JS divergence values for the parameters $\mc$, q, $\chi_{eff}$ and $\widetilde{\Lambda}$.
The top panel of Figure~\ref{JS-div} shows the CDF of the JS-divergences for  \IMRPDTv with \Resum-lmax4 injections, and about
20$\%$ of the population has values greater than 0.1.  These substantial differences persist even if our JS divergence
omits the chirp mass.  In other words, for many events, these two waveform models draw notably different conclusions
about astrophysically pertinent parameters.  

The bottom panel of Figure~\ref{JS-div} provides a scatterplot of the JS divergence versus the expected SNR for each
injection (assessed using a zero-noise realization).  Notably, large JS can occur at any signal amplitude, including
near the detection threshold; however, at large signal amplitude, the JS divergence is never small.  This trend
suggests that systematics can be important for any signal amplitude (i.e., JS $>10^{-2}$), though they will be ubiquitously important for
sources with SNR $\gtrsim 25$.

\begin{figure}[!ht]
\includegraphics[scale=0.35]{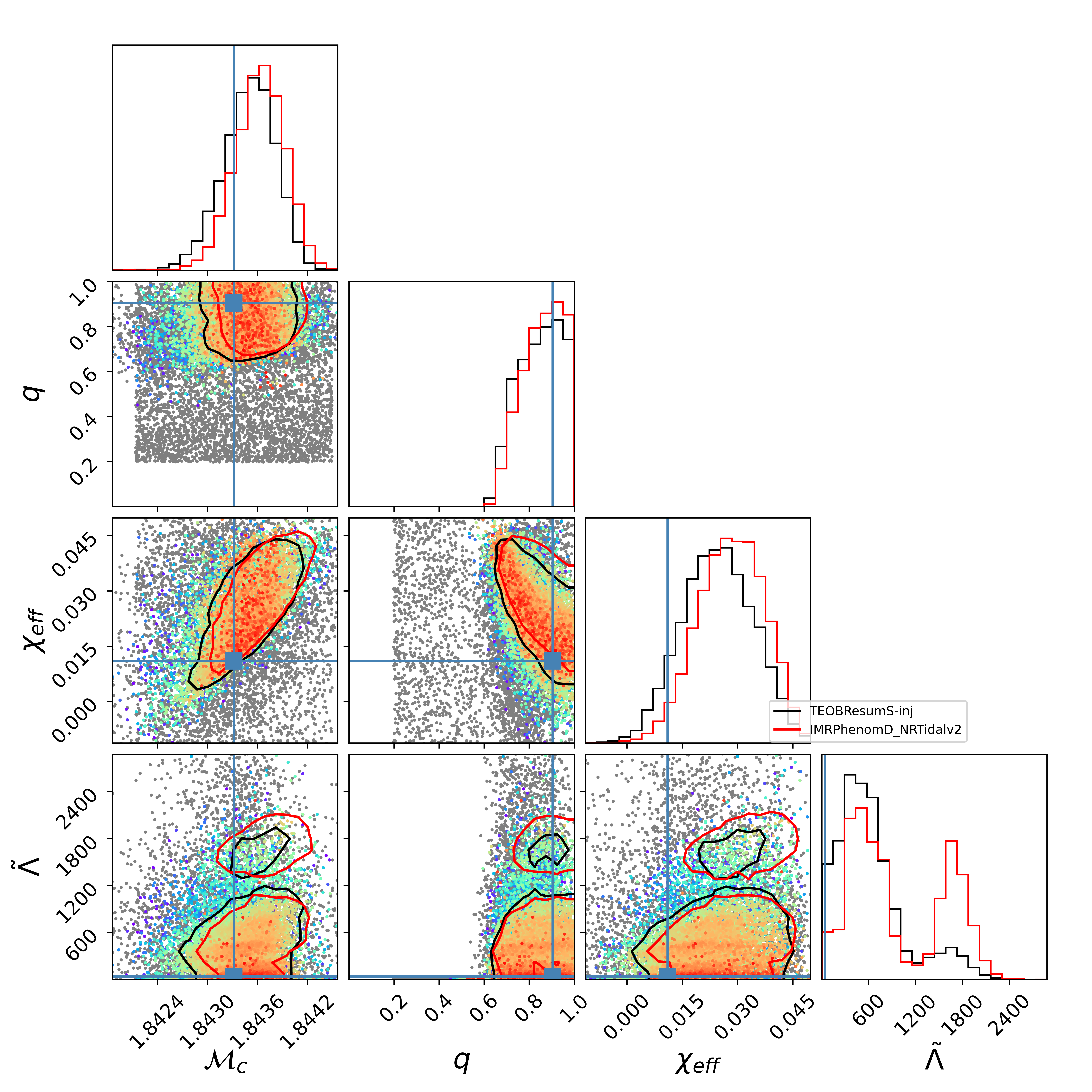}
\includegraphics[scale=0.35]{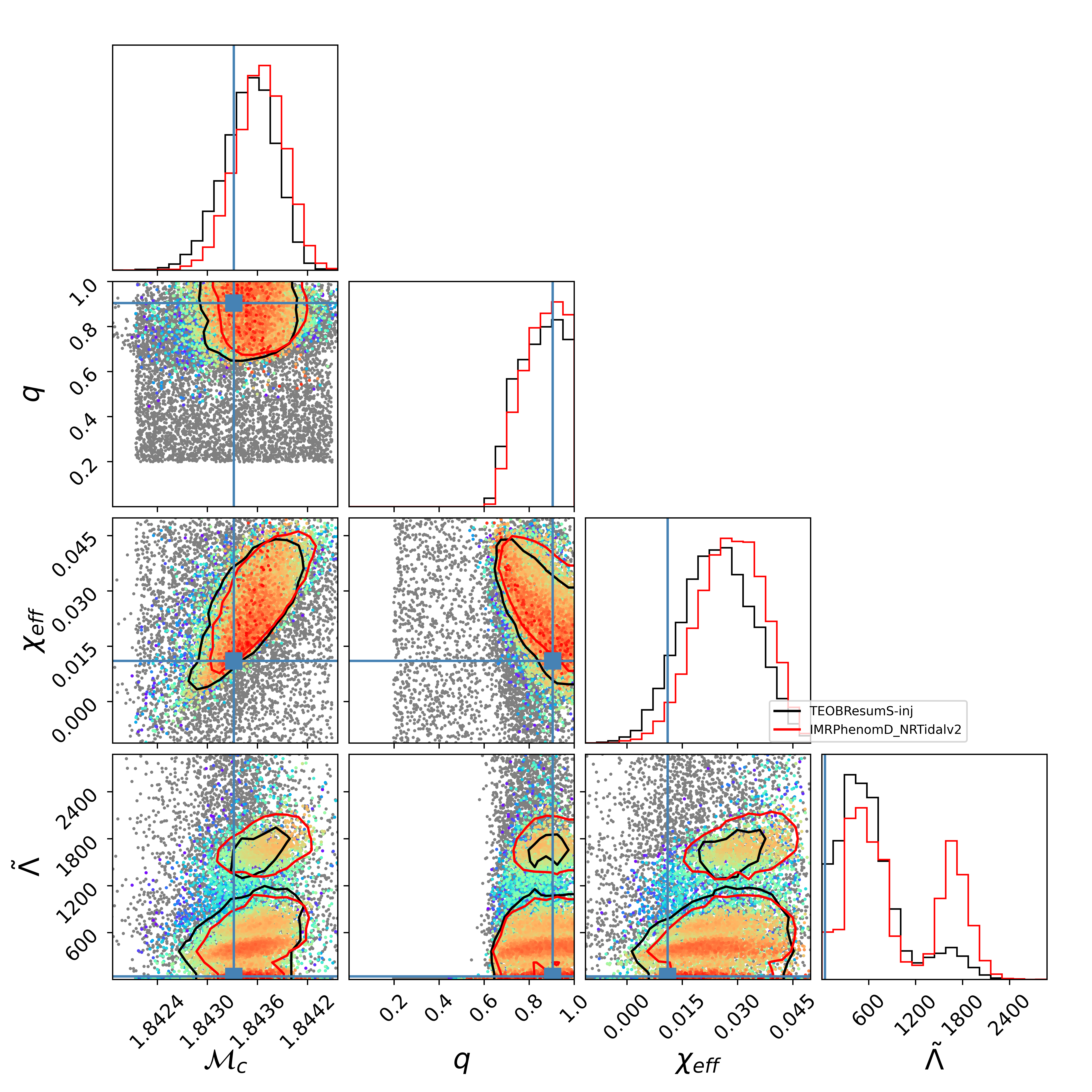}
\caption{Corner plots showing recovered 1D distributions of $\mc$, q, $\chi_{eff}$ and $\widetilde{\Lambda}$ for a \Resum injection 
(cross-hairs indicate true value) analyzed with various waveform models listed in the legend. \emph{Top panel:} Colored grid points from the \Resum run.
\emph{Bottom panel:} Colored grid points from the \IMRPDTv run. The four-parameter JS-divergence value between \Resum and \IMRPDTv for this event is 0.107.}
\label{corner-PE}
\end{figure}

\begin{figure}[!ht]
\includegraphics[scale=0.35]{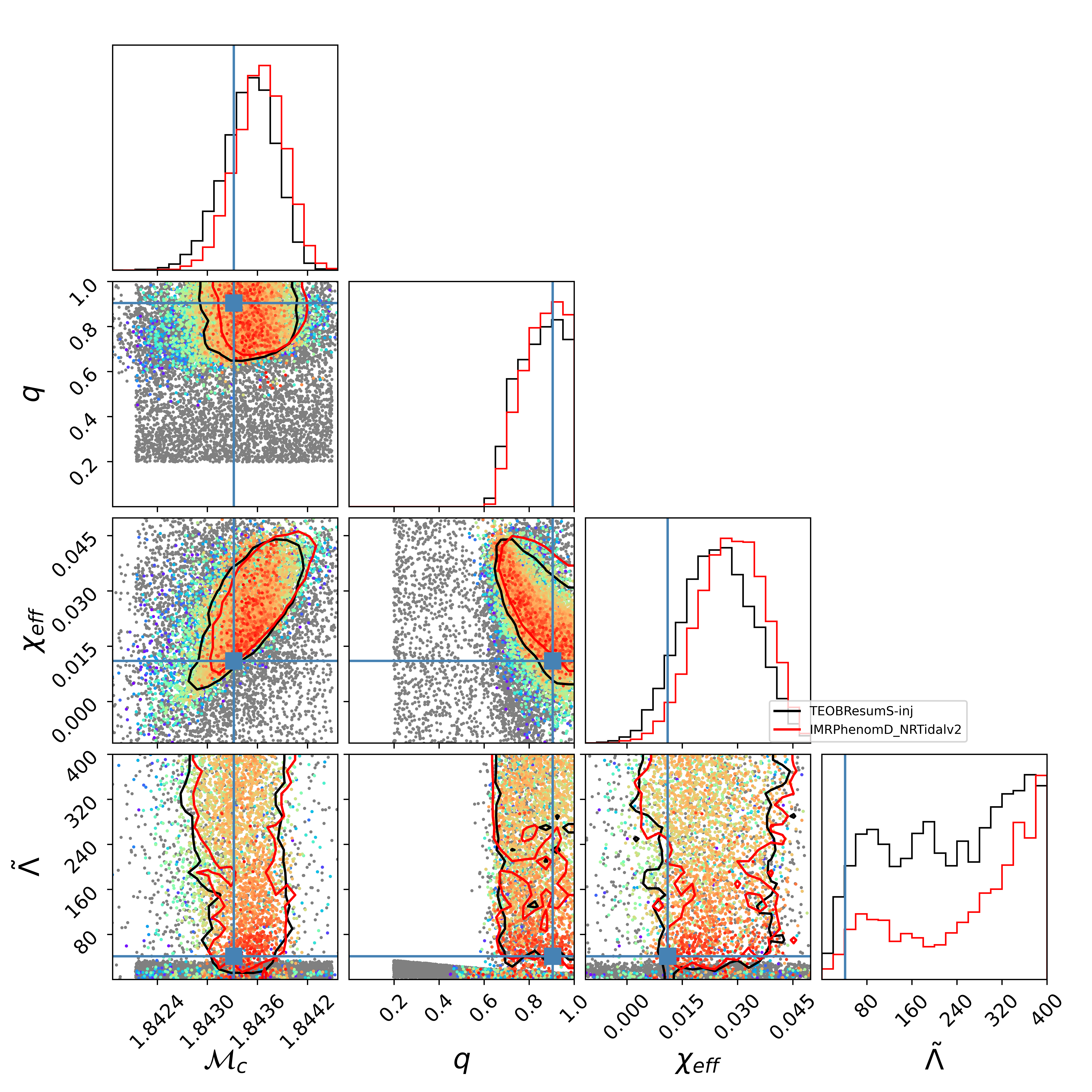}
\includegraphics[scale=0.35]{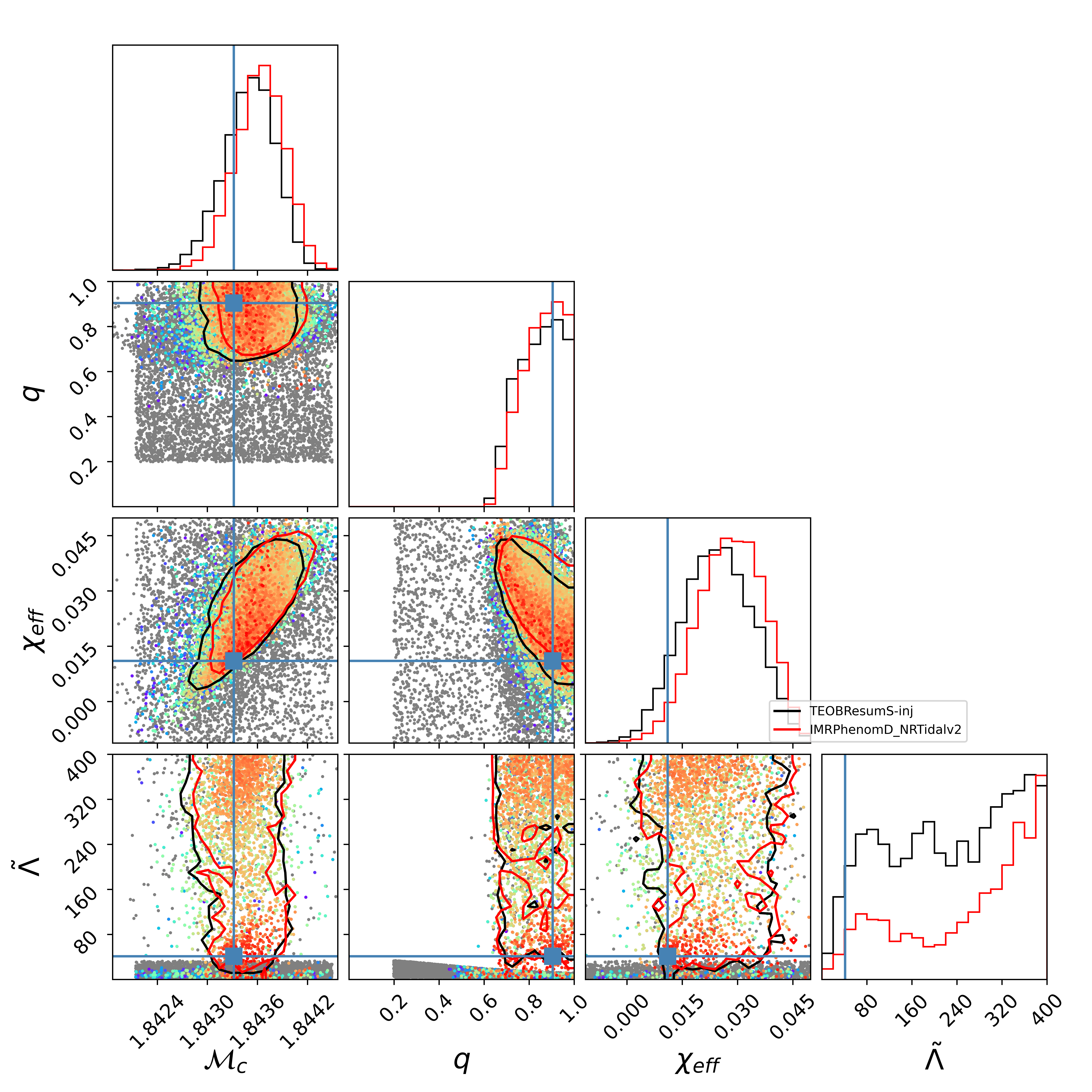}
\caption{Corner plot showing recovered 1D distributions of $\mc$, q, $\chi_{eff}$ and $\widetilde{\Lambda}$ for a \Resum injection 
(cross-hairs indicate true value) analyzed with various waveform models listed in the legend, the same event as in Fig.~\ref{corner-PE} but with a narrower plotting 
range on the tidal deformability parameter. \emph{Top panel:} Colored grid points from the \Resum run.
\emph{Bottom panel:} Colored grid points from the \IMRPDTv run}
\label{corner-PE-lambdanarrow}
\end{figure}

As a concrete example, Figure ~\ref{corner-PE} shows two inferences of the same binary.  This binary has a relatively large total mass
but a modest signal amplitude of 22.31, with a peak marginal log-likelihood of only 200, about half of GW170817's peak.  In both
cases, the tidal deformability is poorly constrained to an interval in excess of what any plausible EoS would
support.  However, within the context of the even broader range of possibilities allowed by our purely phenomenological
tidal parameterization, in which $\Lambda_i$ are uniformly and independently distributed between $[0,5000]$, these two
inferences do differ notably in their conclusions about $\tilde{\Lambda}$.

The example shown in Figure ~\ref{corner-PE} also highlights a critical feature of real PE and the propagation of systematics into
population inference.  Because of the choice of priors adopted for real PE, that inference may (or may not) exhibit
substantial differences between different models.  By contrast, all the EoS models that we consider would predict only
very small $\tilde{\Lambda}$ in this mass range. As a result, the JS divergence (though a useful guide to identifying
coarse model differences) doesn't reflect our prior knowledge about what physics can support at these high masses.
Figure~\ref{corner-PE-lambdanarrow} shows the same analysis as Figure ~\ref{corner-PE}, now focused on a smaller range
of $\tilde{\Lambda}$.  In this focused range, the two analyses exhibit even larger differences in their marginal
posterior for $\tilde{\Lambda}$ and in the likelihood evaluations used to draw that conclusion (see, e.g., the
differences in red points between the left and right figure).  This focused representation highlights the fact that
the impact of systematics is highly uneven across the posterior, and small differences globally may disguise larger
differences on smaller but physically critical scales.

\subsection{Population inference: An extreme scenario}

\begin{figure}
\includegraphics[scale=0.5]{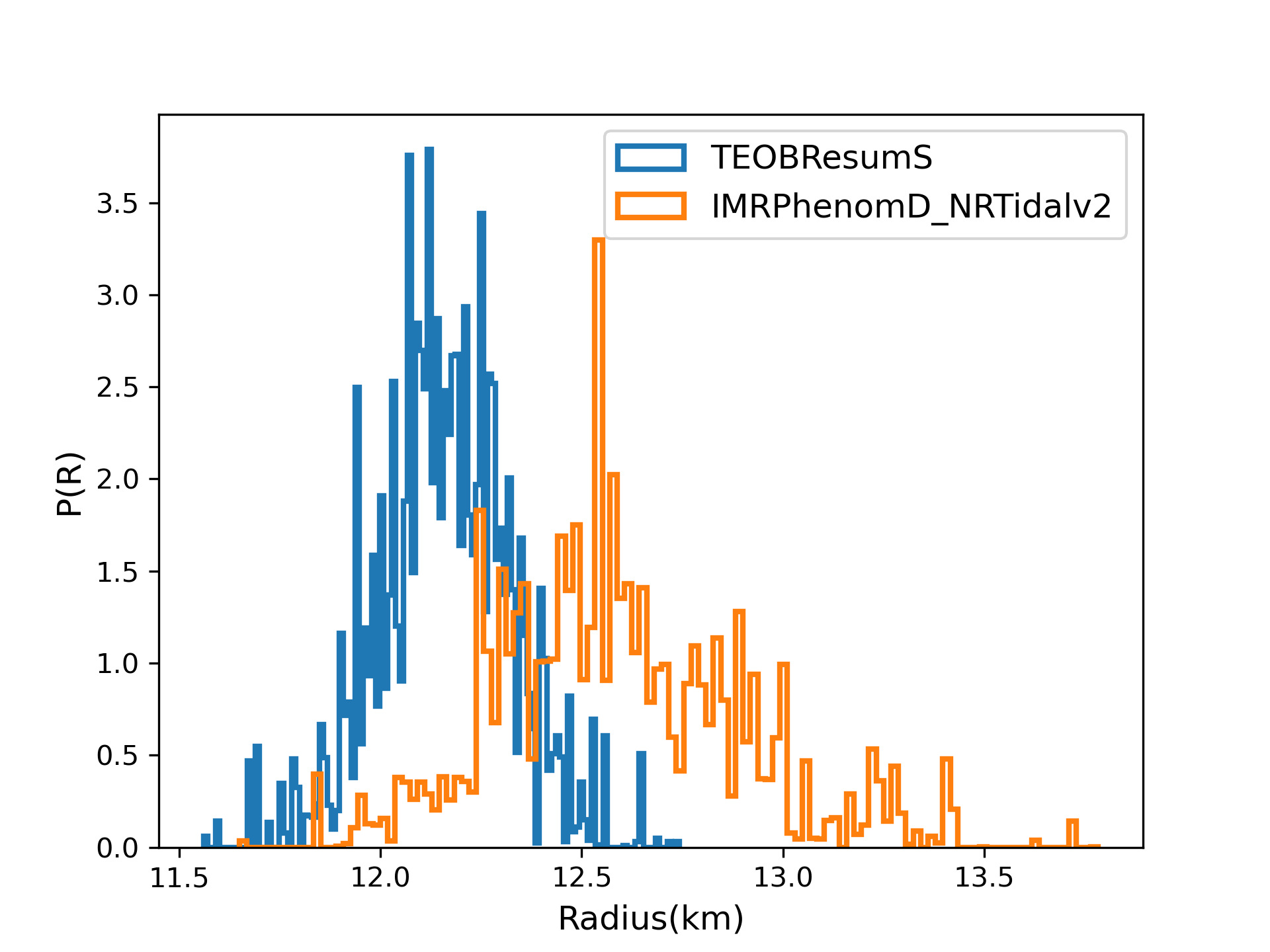}
\includegraphics[scale=0.5]{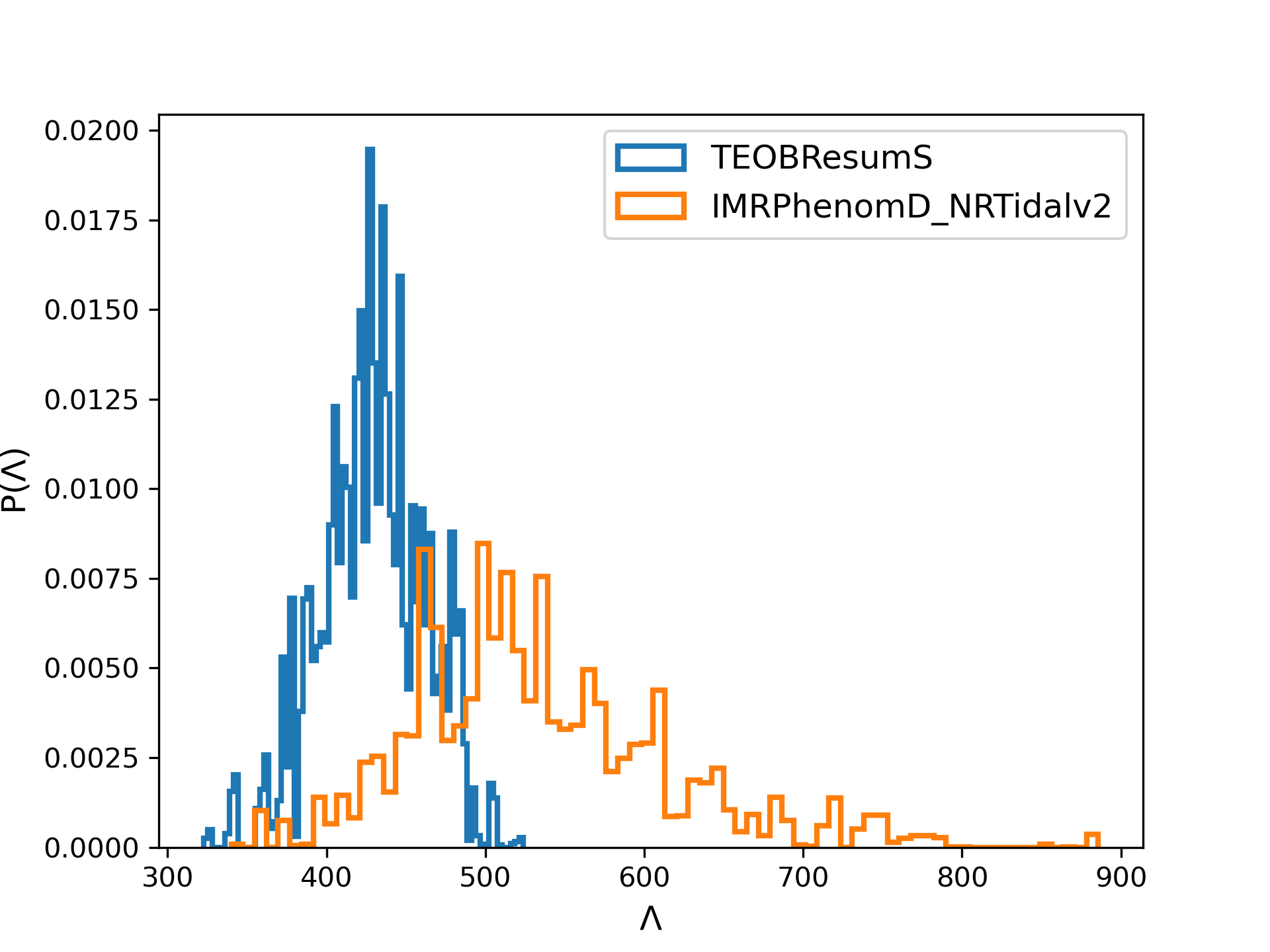}
\caption{\emph{Top panel}: Posterior distribution for the radius of a typical 1.4$\Msun$ neutron star for the equations-of-state realizations from the joint population and EoS inference.
\emph{Bottom panel}: Posterior distribution of the dimensionless tidal deformability of a 1.4$\Msun$ neutron star for the equations-of-state realizations from the joint population and EoS inference. Legend indicates the waveform model that was used in the PE for the respective population analyses.}
\label{radius_plot}
\end{figure}

\begin{figure}
\includegraphics[scale=0.5]{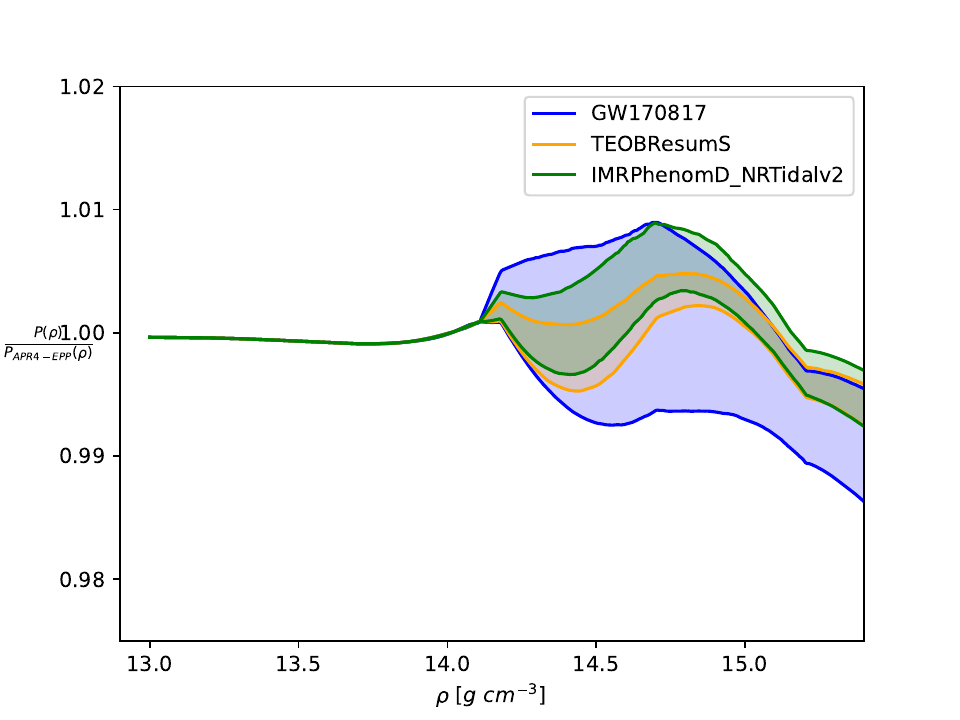}
\caption{Pressure-density plot comparing population analyses results from the different parameter estimation results 
performed for the same \Resum-lmax4 injection waveform model indicated in the legend. Specifically, it plots 
the ratio of pressure estimates from the recovered EoSs to that of pressure estimates from the injected equation-of-state, 
APR4. The shaded region indicates the 90$\%$ CI of the respective analyses.}
\label{P_rho}
\end{figure}

For the joint population and equation-of-state inference, we selected ten events whose JS-divergences values from PE were highest between \Resum and \IMRPDTv.
Two sets of such analyses were performed- one for each of the waveform models assumed in PE. Ensuring convergence of the analyses, we then
 looked at the radius and tidal deformability distribution of a 1.4$\Msun$ neutron star for the various EoSs inferred from the population analyses. 
Although Fig. ~\ref{P_rho} shows that the information from both analyses differs very little, they are important enough to exhibit systematics for
other parameters. Figure ~\ref{radius_plot} shows a significant difference between the inferred 
radius and tidal deformability from the two analyses. \Resum consistently infers a smaller radius and tidal deformability compared to \IMRPDTv in this scenario. 
The \Resum model predicts a median NS radius  of $\MedianOf{R} =12.16$ km with a standard deviation of 0.17 km while for \IMRPDTv
they are 12.55 km and 0.31 km, respectively, meaning the inferred radius differs by 2$\sigma$ in one case. The tidal deformability distributions also show a similar trend with 
$\MedianOf{\tilde{\Lambda}}$=429.28, $\sigma$=32.35 for \Resum and $\MedianOf{\tilde{\Lambda}}=519.428$, $\sigma$=81.69 for \IMRPDTv. The $R_{1.4}$ and $\Lambda_{1.4}$ distribution for \IMRPDTv also have a wider 
spread compared to \Resum.

For both of our analyses of the same fiducial ten-event sample, we also calculate the rest mass density versus pressure trends from the inferred adiabatic index coefficients. Figure ~\ref{P_rho} plots the ratio of pressure
estimates from the recovered EoSs to that of pressure estimates from a fiducial reference EoS (APR).  We limit ourselves
to densities less than 3$\times 10^{15} g cm^{-3}$.  For comparison, we also show the results inferred from the original
LVK interpretation of GW170817~\cite{170817_EoS_data}, derived from hyperparameter samples using the same spectral EoS parameterization and
implementation. [For efficiency, these samples were also used to initialize our population inference calculations.]  
As expected, the GW170817 EoS posterior encompasses the inferences derived using both waveform models.  Too, inferences
derived using the two waveform models differ modestly at moderate density, consistent with the modest differences
seen for fiducial NS radius in Figure~\ref{radius_plot}.



As demonstrated qualitatively in Figure \ref{JS-div} and Figure \ref{JS-Mc} in the Appendix, the ten events identified here are not exceptional in signal
amplitude nor source parameters. Many sources have only modestly lower JS divergence than the events identified
here.  Therefore, we expect that while this analysis highlights a relatively unlikely scenario, it remains suggestive of
a broader issue: a long tail of events with large systematics, where those systematics cannot be identified save by
analysis with multiple waveform families.  When incorporated into a population study, these events will
impact our global conclusions about source parameters in proportion to the fraction of such events included.


\subsection{Population analysis: Representative samples}

\begin{figure}[!ht]
\includegraphics[scale=0.5]{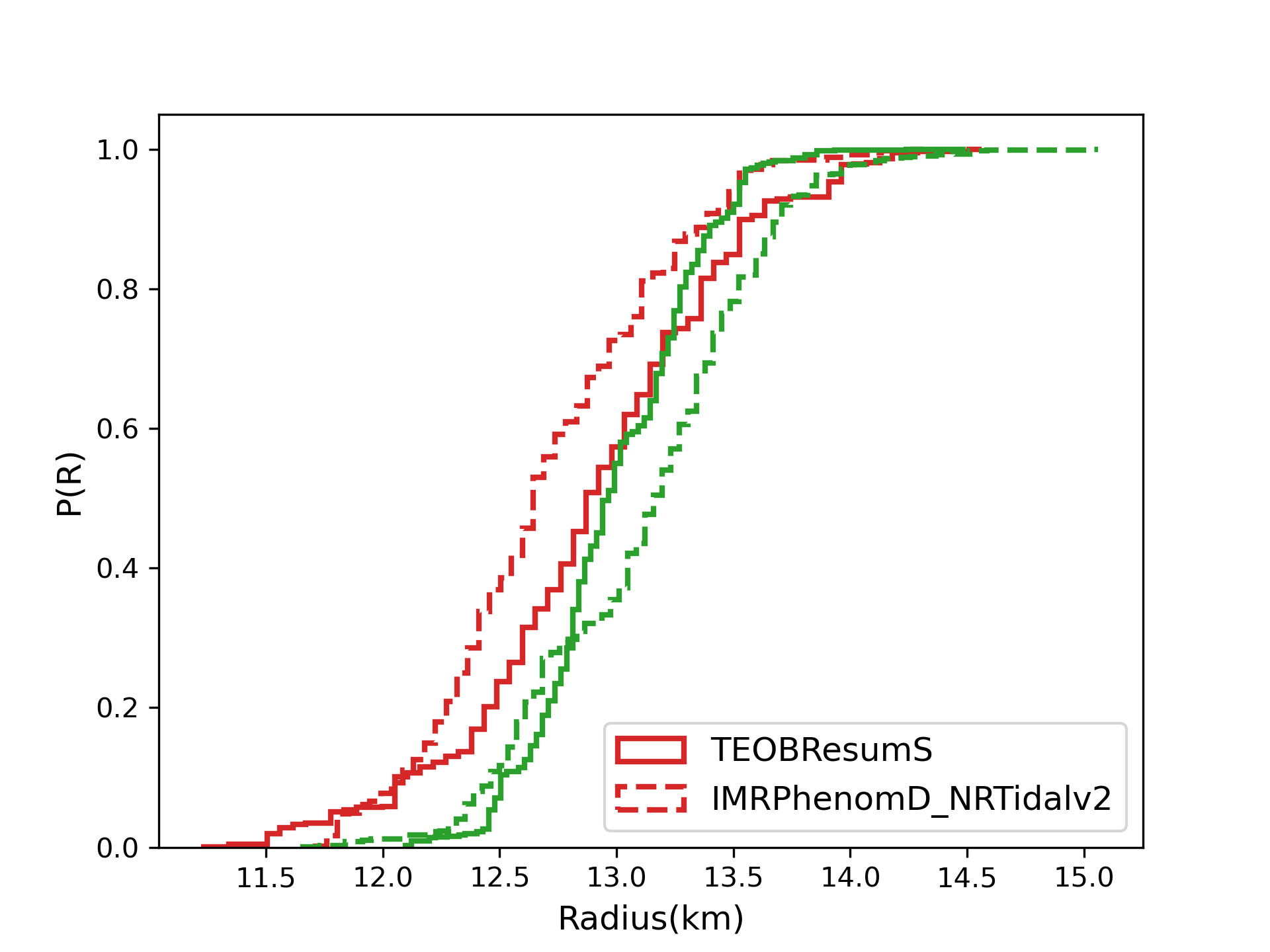}
\includegraphics[scale=0.5]{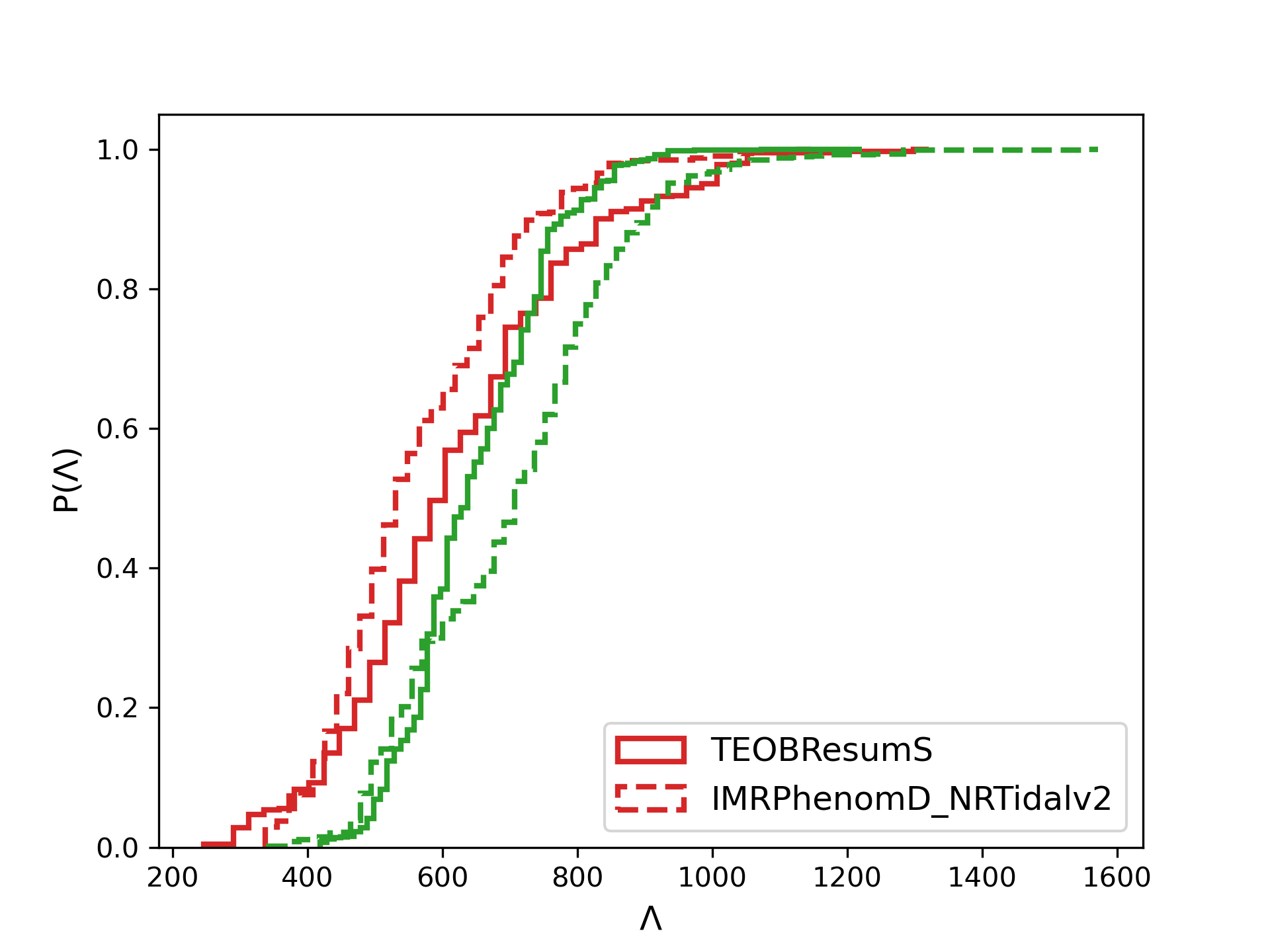}
\caption{\emph{Top panel}: Histogram of the radius of a 1.4$\Msun$ neutron star for the equations-of-state realizations from the joint population and EoS inference.
\emph{Bottom panel}: Histogram of the dimensionless tidal deformability of a 1.4$\Msun$ neutron star for the equations-of-state realizations from the joint population and EoS inference. Legend indicates the waveform model used in the PE for the respective population analyses.}
\label{radius_plot_random}
\end{figure}

To compare against our choice of picking events with large differences in PE between waveform models, we also perform similar population analyses 
but with two  random choices of 10 events from the population, i.e., without taking the JS divergences value into
consideration. Figure ~\ref{radius_plot_random} shows the cumulative posterior probability for the 
radius and tidal deformability distribution of a 1.4$\Msun$ star similar to Fig. ~\ref{radius_plot}.   The red and green colors indicate results derived from the two sets of 10 events from our population.  The solid lines
indicate analyses derived using the same waveform model used to generate our synthetic signals (here, \Resum).  These two results are
very similar, with modest differences in the posterior distribution due to the specific choice of ten events used in the
analysis.  Notably,  the median of both distributions is nearly identical: for \Resum, the median difference between the
two realizations is only $\Delta R_{0.5}=0.05\unit{km}$.
By contrast, the dashed lines show the population results derived if each event is analysed with a different waveform
model family (here, \IMRPDTv).  Though neither analysis exhibits as extreme systematic bias as the cherry-picked
analysis described above, these two analyses do demonstrate that even a small-sample analysis generally incorporates
a significant unmodeled systematic error.  For example, the medians of the two dotted lines differ visibly and in a statistically
significant way from one another and from the (common) median of the \Resum analyses shown with solid lines.   To be
concrete, the median difference between the two \IMRPDTv realizations is $0.5\unit{km}$ -- ten times larger than the
difference for \Resum -- and are roughly $\pm 0.25$ from the common median seen for \Resum.   Though
 smaller than the width of each posterior distribution, which accounts only for statistical uncertainty, this comparison
 demonstrates that waveform systematics introduce excess noise and that a few-event analysis can resolve its
 effect.



\begin{figure}[!ht]
\includegraphics[scale=0.5]{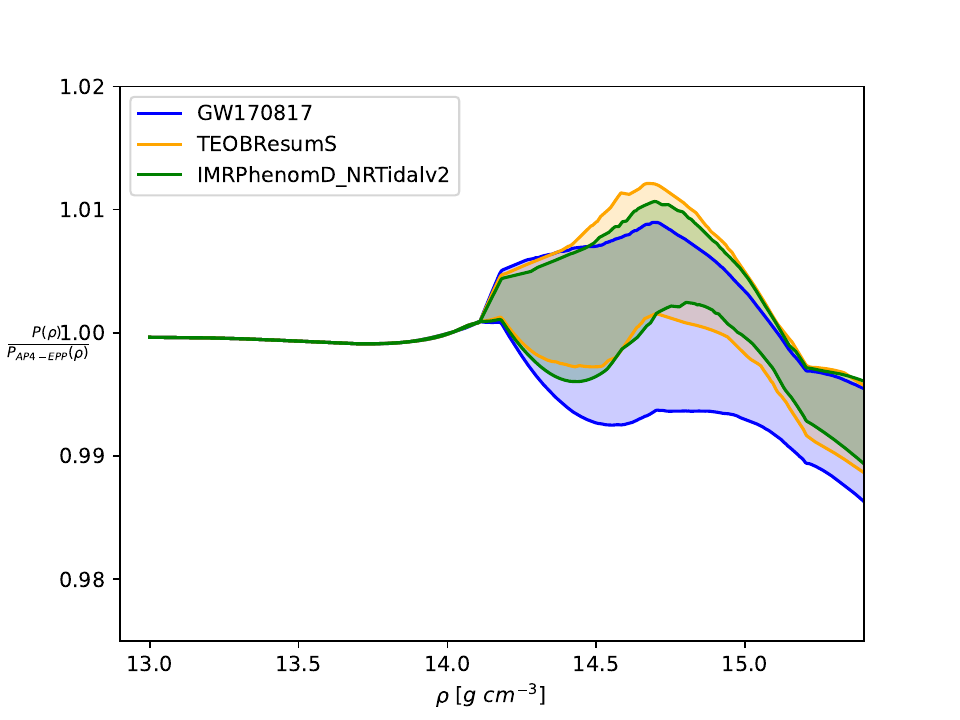}
\caption{Pressure-density plot comparing population analyses results from the different parameter estimation results 
performed for the same \Resum-lmax4 injection, waveform model indicated in the legend. Specifically, it plots 
the ratio of pressure estimates from the recovered EoSs to that of pressure estimates from the injected equation-of-state, 
APR4. The shaded region indicates the 90$\%$ CI of the respective analyses.}
\label{P_rho_random}
\end{figure}


\subsection{Discussion and ansatz}

From the above analyses, we conclude that substantial waveform systematic uncertainties exist across the model space,
exemplified by ubiquitous differences in marginal likelihoods and pertinent posteriors.  Further, based on a census of
our cross-waveform differences, we further anticipate that (as in prior work) these differences depend strongly on each
specific event's intrinsic (and potentially extrinsic) parameters.   As a result, waveform systematics have two
qualitatively different impacts on our inferences of the NS radius.  First, they introduce a bias in our inferred
radius, to the extent that each event on average contributes to a mean offset in $R_{1.4}$.   By construction, this bias
reflects the population average of the (unknown) systematic bias $\Delta R$ over the merging NS population.  Second,
because different events in the NS population produce different realizations of the systematic offset $\Delta R$, for
any specific set of events chosen, we will infer different answers for the radius.  Asymptotically and assuming that
second moments of $\Delta R$ exist, these
realization-specific differences should converge to  $\sigma_{\Delta R}/\sqrt{N}$, where $\sigma_R$ is the
population-averaged standard deviation of the bias $\Delta R$.

Our numerical examples above are both consistent with this picture, and suggest that the population-averaged bias $\E{\Delta R}$ is small, compared to the excess dispersion
from the standard deviation $\sigma_{\Delta R}$ of $\Delta R$ over our population.
Our observation differs from multiple previous attempts to identify systematic biases from similar end-to-end
population inference calculations.  In brief and as discussed below, previous investigations usually examined a single
sequence of BNS mergers, interpreting their $R(N)$ trajectory and particularly its endpoint as a measure of the average
bias $\E{\Delta R}$.  By contrast, our calculations suggest that {hundreds of}  BNS mergers should be analyzed
self-consistently to resolve the systematic realization dependent error ($\sigma_{\Delta R}/\sqrt{N}$) from the
expected population-averaged waveform bias $\E{\Delta R}$, for the populations considered here.

Finally, the above ansatz suggests that the key parameters $\E{\Delta R}$ and $\sigma_R$ can be  measured from
the underlying parameter inference and propagated into population studies.  As discussed earlier, in principle the pertinent differences
which propagate into population-inferred EOS constraints arise in
the underlying parameters inferences only  from changes at small $\Lambda$.  In practice and as a first approximation to
characterize trends in systematics across the source binary parameter space, we will instead measure changes in the median
value of $\Lambda$ inferred for each event under the fiducial parameter inference prior.   While this estimates for
$\Lambda$ are 
derived using an overly conservative and uninformative prior for the binary components masses and tides, we anticipate
that it will nonetheless capture the dominant trend -- rapid  variation in $\Delta R$ across the binary parameter space,
which over the population nearly averages to zero but with large variance.

\subsection{Quantifying dispersion in $R_{1.4}$ and $\Lambda_{1.4}$}


To understand the dispersion of the recovered $R_{1.4}$ and $\Lambda_{1.4}$ posterior distributions for different sub-sets of the injected population, we use a simple model of estimating the error of a quantity when multiple observations/data is stacked. For this, we choose to compute this quantity for the radius of a 1.4$\Msun$ neutron star from parameter estimation samples, using Eq.18 in Zhao $\&$ Lattimer~\cite{Zhao:2018nyf}, defined below:
\begin{eqnarray}
R_{1.4} \simeq (11.5 \pm 0.3) \frac{\mathcal{M}}{\Msun} \Bigl(\frac{\widetilde{\Lambda}}{800}\Bigl)^{1/6}
\label{R_14_calc}
\end{eqnarray}
Applying this expression to the median value of $\tilde{\Lambda}$ derived using the two different waveform
models \Resum and \IMRPDTv, we can furthermore for each event estimate a per-event estimate of difference in radius
solely due to systematic waveform differences:
\begin{eqnarray}
\Delta R_{1.4} = R_{1.4}^{\Resum} - R_{1.4}^{\IMRPDTv}
\label{diff}
\end{eqnarray}
The quantity $\Delta R_{1.4}$ varies substantially on a per-event basis, in part due to changes in systematic biases
over the parameter space.    For any subset of $N$ events, we compute the
sample mean $\overline{\Delta R} = \sum \Delta R_{1.4}/N$ and sample standard deviation $s_{\Delta R}$.  For a
sufficiently large sample, these two quantities should converge to the population-averaged mean $\mu_{\Delta R}$ and
standard deviation $\sigma_{\Delta R}$ of $\Delta R$.




              

\begin{figure}
\includegraphics[scale=0.5]{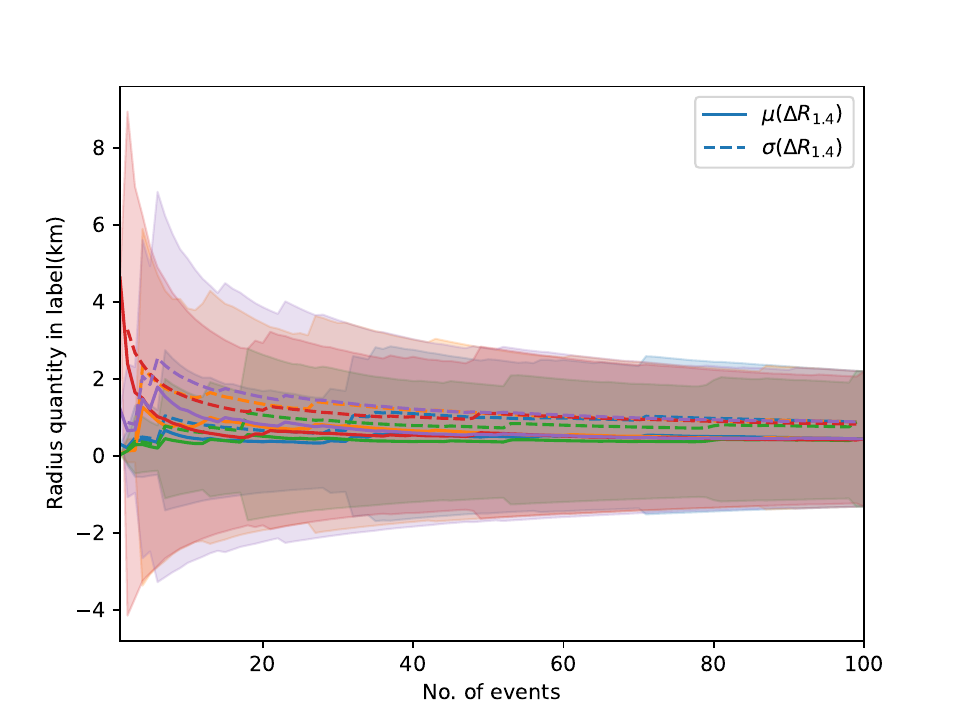}
\includegraphics[scale=0.5]{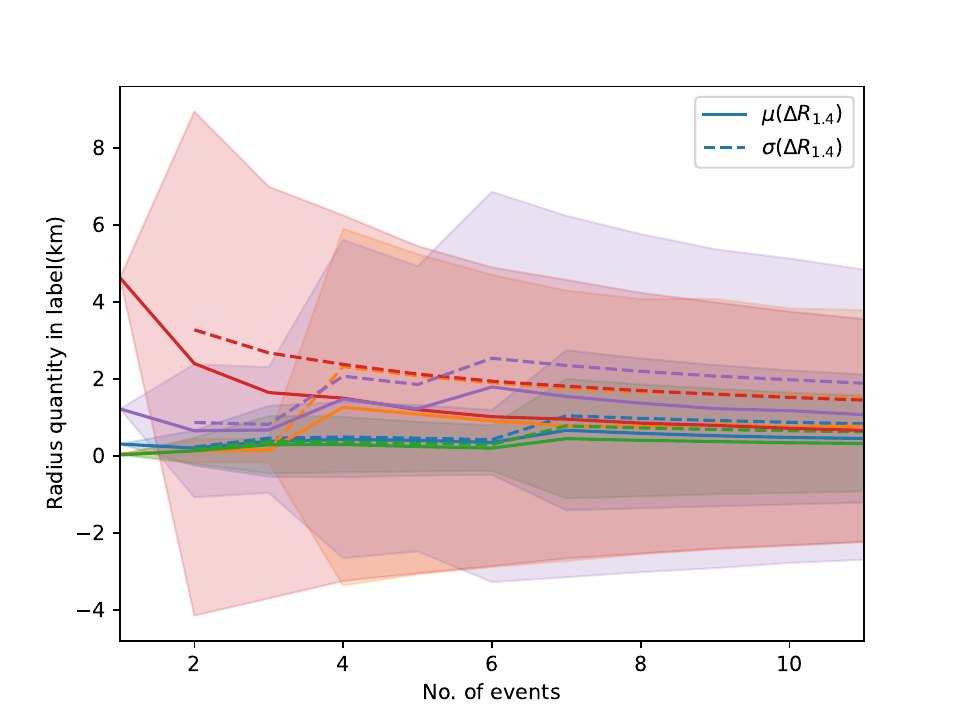}
\caption{\emph{Top panel}: Plot showing how different stacking realizations of the 100-event simulated population converge to the same $\Delta R$ and its corresponding 
standard deviation after consideration of a certain high number of events. The shaded region shows the $\pm 2\sigma$ region of $\mu(\Delta R)$ for each realization.
\emph{Bottom panel}: Same plot as the top panel with the x-axis zoomed in to show the behavior for N=10 events.}
\label{fig:stats_sys}
\end{figure}

Figure~\ref{fig:stats_sys} shows the sample mean (solid curves) and sample standard deviation  (dashed lines) of $\Delta R$ derived from
different sample realizations, with the $x$ axis indicating the number of samples used and the $y$ axis measuring these
two quantities (in km).  Going from left to right, each solid-colored line calculates the sample mean using the first
$N$ events of a specific ordering of our 100 synthetic events.  The different colors reflect five different choices of
ordering.  

%
At the right side of the graph, all estimates converge to an estimate of the true  population-averaged mean $\mu_{\Delta
  R}\simeq 0.44\unit{km}$ and
true  population-averaged standard deviation $\sigma_{\Delta R} \simeq  0.88 \unit{km}$.  [In these estimates, we have not attempted any
  corrections to account for finite sample size.]
Using these two parameters, we anticipate that for sufficiently large $N$, the sample mean $\overline{\Delta R_{1.4}}$ should have mean
$\mu_{\Delta R}$ and standard deviation  $\sigma_{\Delta R}/\sqrt{N}$. The shaded region shows the $\pm 2\sigma$ region of the sample mean, computed 
for each realization.

%
We emphasize that the differences characterized  here with a  random variable  arise due to waveform
systematics whose impact changes across the binary parameter space.

In terms of these empirical measurements characterizing systematic uncertainties, we anticipate that for an analysis
comparable to 
Figure~\ref{radius_plot} -- using $N=10$ events -- we would expect to see inferences derived using these two different
waveform models to have medians that vary roughly according to a normal distribution with standard deviation comparable to
$\sigma_{\Delta R_{1.4}}/\sqrt{N} \simeq 0.27\unit{km}$ and systematic average error of order $0.4\unit{km}$.   Notably,
the difference in median is  in good qualitative agreement with the features seen in Figure~\ref{radius_plot}.
Furthermore, for $N=10$ events we find $\sigma_{\Delta R_{1.4}}/\sqrt{N} $ is comparable to $\mu_{\Delta R}$.  As a
result, we expect that with only tens of events, the posterior mean may very significantly solely due to the specific
choice of 10 events selected, consistent with the findings shown in Figure~\ref{radius_plot}.  Only with a large sample
size can we disentangle the relative contributions of mean and random systematic error.

\section{Discussion}
\label{sec:discussion}

There have been multiple approaches to understanding the impact of GW modeling systematics on the inferred EoS derived
from multiple GW observations.
 Kunert and collaborators~\cite{Kunert:2021hgm}  looked at the effect of waveform systematics on EoS inference when combining information from multiple 
 detections for EoS inference from a discrete a priori grid, fixing a uniform NS population model.
 Their study used a fixed set of 38 synthetically-detected events, drawn from a uniformly distribted sample from their population
 model and volume, using an SNR threshold of 7.
They performed population inference for the EOS, as characterized by $R_{1.4}$, for populations of 1 to 38 events,
obtained by consecutively increasing their sample size by one.  They repeated this approach 1000 different times, using
different orderings of their 38 events (see, e.g., their Fig 2).   As a result, for $N$ sufficiently small compared to
their total sample set, this study did use multiple randomly chosen synthetic universes from a larger
sample~\cite{Wysocki:2020myz}; however, their overall final results for systematic biases  relies on a single set of 38 events.
Like our work, they found that the result of their population inference depended on the choice of waveform model used to
interpret the events, with shifts of up to 400m depending on the waveform used.   They demonstrated that a large sample
size (here, their 38 events) was needed before these systematics were comparable to the posterior's width.
[The posterior incorporates only statistical uncertainty, not systematic uncertainty.]
However, due to the limitations of a single data realization, unlike our work and  some others ~\cite{Wysocki:2020myz}
Kunert et al. does not characterize  the increased dispersion that occurs due to waveform systematics, or discuss its
impact on their final $N=38$ fixed-sample-size analysis.
Rather than
producing a single unique bias, our study shows that (as should be expected) systematics also contributes to enhanced dispersion
of the posterior, allowing for multiple realizations.   In fact, the result of Kunert et al. for $N=38$ is consistent with the same
scale of random systematic dispersion suggested in our analysis.  Their asymptotic result, which relies on a single
realization, could reflect the true mean or instead dispersion.

Kunert and collaborators~\cite{Kunert:2021hgm} also demonstrate that even though the net effects of waveform systematics
for joint EoS inference
are significant, on a per-event basis these systematic biases are relatively small compared to the per-event source
parameter posterior.   This conclusion has been corroborated by other studies like  Gamba et al. ~\cite{Gamba:2020wgg},
who argue that on a per-event basis systematic uncertainties will only be significant for sources with  SNR$\geq$80.  In
our investigation, however, the maximum SNR  of an event is $\sim$ 50. 

An article by Chatziioannou~\cite{2022PhRvD.105h4021C} uses physics-based analysis derived from the underlying
gravitational wave phase to investigate different effects that could lead to uncertainties in the neutron star radius
measurements.  Her semianalytic analysis suggests that, with the current detectors and observations, the
statistical uncertainties are a bigger factor than systematic uncertainties in determining the equation-of-state.
These uncertainties start becoming comparable once a statistical error of 0.5-1 km is reached at the 90$\%$
credible level with A+ sensitivity. Pratten et al. ~\cite{Pratten:2021pro} looks at the effect of exclusion of dynamical tides on the EoS reconstruction on a 
population of BNS, but one that is limited to lower-mass centered around 1.33$\Msun$. Although they see $\mathcal{O}$(1 km) 
systematic differences in the recovered radius with \texttt{TaylorF2}, it is possible that using waveform models with 
higher-order models included could be responsible for better constraints and smaller error bars in our case. 

Neither of the studies mentioned above performs a joint inference of the population properties of BNS sources along with the equation of state, 
which could bias the understanding of the inferred EoS, as demonstrated in \cite{Wysocki:2020myz, Golomb:2021tll, Ray:2022hzg}. Irrespective 
of that, there is a surprising consensus that the systematics are comparable to statistical effects in the order of 10 events, seen by us, Kunert et al. ~\cite{Kunert:2021hgm}
 and Chatziioannou~\cite{2022PhRvD.105h4021C}.


\section{Conclusions}
\label{sec:conclude}

In this paper, we investigated the impact of waveform systematics on a joint population and equation-of-state inference for a fiducial population of
moderate-amplitude binary neutron star sources. Similar to ~\cite{Yelikar:2024wzm}, firstly, we use JS-divergences to quantify differences in the parameter estimation 
analysis, with a figure to illustrate a single extreme example from our large sample.
Further,
we use these posterior samples to perform multiple joint hierarchical inferences of the population and EoS properties of the
fiducial systems by selecting different synthetic universes of ten of these events, following previous work \cite{Wysocki:2020myz}. We found that
when the ten events are picked such that they have the worst JS divergence between the two waveform models, then the
inferred EOS has a large systematic difference compared to the statistical error. We
also performed a similar population analysis, choosing events at random and observing a more modest and random systematic
difference between the inferred EOS.  In other words,  the systematically different events introduce downstream uncertainties
into any calculations into which they are incorporated, increasing the systematic dispersion of EOS inferences.

To the best of our knowledge, this is one of the first works to propagate systematic error in end-to-end BNS EOS
inference, accounting for multiple data realizations and using a sufficiently large sample, allowing us to probe better the
long tail of rare but large differences between different waveform models in the NS parameter space.  In the future, we plan to use this for future detections of BNS sources as well as make 
improvements in our current setup, such as combining information from neutron star-black hole (NSBH) sources. 



\begin{acknowledgements}

The authors would like to thank Anarya Ray for helpful comments. ROS  gratefully acknowledges support from NSF awards NSF PHY-1912632, PHY-2012057,  AST-1909534, and the Simons Foundation.
AY acknowledges support from NSF PHY-2012057 grant. This material is based upon work supported by NSF’s LIGO Laboratory 
    which is a major facility fully funded by the
    National Science Foundation.
This research has made use of data,
    software and/or web tools obtained from the Gravitational Wave 
    Open Science Center,
    a service of LIGO Laboratory,
    the LIGO Scientific Collaboration and the Virgo Collaboration.
LIGO Laboratory and Advanced LIGO are funded by the 
    United States National Science Foundation (NSF) as
    well as the Science and Technology Facilities Council (STFC) 
    of the United Kingdom,
    the Max-Planck-Society (MPS), 
    and the State of Niedersachsen/Germany 
    for support of the construction of Advanced LIGO 
    and construction and operation of the GEO600 detector.
Additional support for Advanced LIGO was provided by the 
    Australian Research Council.
Virgo is funded through the European Gravitational Observatory (EGO),
    by the French Centre National de Recherche Scientifique (CNRS),
    the Italian Istituto Nazionale di Fisica Nucleare (INFN),
    and the Dutch Nikhef,
    with contributions by institutions from Belgium, Germany, Greece, Hungary,
    Ireland, Japan, Monaco, Poland, Portugal, Spain.
The authors are grateful for computational resources provided by the 
    LIGO Laboratory and supported by National Science Foundation Grants
    PHY-0757058, PHY-0823459, PHY-1626190 and PHY-2110594.
    We acknowledge the use of IUCAA LDG cluster Sarathi for the 
computational/numerical work. 

\end{acknowledgements}

\appendix

\section{Chirp mass and JS divergence}
\label{ap:Mc}

Similar to Fig.~\ref{JS-div} (bottom panel), we plotted the chirp-mass of the system against the JS divergence values of parameter inference 
samples of the two waveform models, and do no observe any correlation.

\begin{figure}
\includegraphics[scale=0.5]{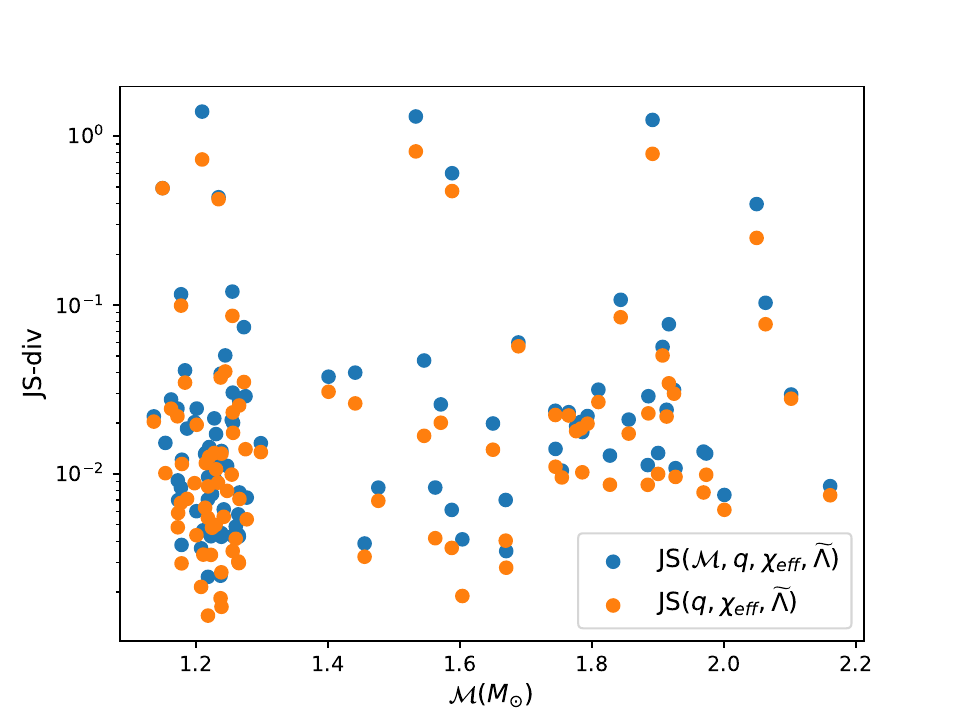}
\caption{Scatter plot showing how the chirp mass of the injections from the population relates to the JS divergence value between 
\Resum and \IMRPDTv samples for the parameters noted in the legend.}
\label{JS-Mc}
\end{figure}

\bibliography{references,paperexport,LIGO-publications,EoS-constraints-papers}

\end{document}